\documentclass{aa}
\usepackage{natbib}
\usepackage{graphicx}
\usepackage{amssymb}

\bibpunct{(}{)}{;}{a}{}{,}

\begin{document}

\title{Morphology of galaxies  with quiescent recent assembly history in a $\Lambda$-CDM universe}

\author{S. E. Pedrosa\inst{1,2}
        \and P. B. Tissera \inst{1,2,3}
        \and M. E. De Rossi \inst{1,2,4}
        }

\offprints{S. E. Pedrosa}

\institute{Consejo Nacional de Investigaciones  Cient\'{\i}ficas y T\'ecnicas, CONICET, Argentina.\\
\email{supe@iafe.uba.ar}
\and Instituto de Astronom\'{\i}a y F\'{\i}sica del Espacio, Casilla de Correos 67, Suc. 28, 1428, Ciudad Aut\'onoma de Buenos Aires, Argentina.
\and Departamento de Ciencias Fisicas, Universidad Andres Bello, Av. Republica 220, Santiago, Chile.\\
\email{patricia@iafe.uba.ar}
\and Facultad de Ciencias Exactas y Naturales, Universidad de Buenos Aires, Ciudad Aut\'onoma de Buenos Aires, Argentina.\\
\email{derossi@iafe.uba.ar}
}

\date{Accepted}

\abstract {The standard disc formation scenario postulates that  disc forms as the gas cools and flows into the centre of the dark matter halo, conserving the specific angular momentum. Major mergers have been shown to be able to destroy or highly perturb the disc components. More recently, the alignment of the material that is accreted to form the galaxy has been pointed out as a key ingredient to determine galaxy morphology. However,  in a hierarchical scenario galaxy formation is a complex process that combines these processes and others in a non-linear way so that the origin of galaxy morphology remains to be fully understood. }
{We aim at exploring the differences in the formation histories of galaxies with a variety of morphology, but quite recent merger histories, to identify which  mechanisms are playing a major role. 
We analyse when minor mergers can be considered relevant to determine galaxy morphology.
We also study the specific angular momentum content of the disc and central spheroidal components separately. 
}
{We used cosmological hydrodynamical simulations  that include an effective, physically motivated supernova feedback that is able to regulate the star formation in haloes of different masses. 
We analysed the morphology and formation history of a sample of 15 galaxies of a cosmological simulation. We performed a spheroid-disc decomposition of the selected galaxies and their progenitor
systems. The angular momentum orientation of the merging systems as well as their relative masses were estimated to analyse the role played by orientation and by minor mergers in the determination of the morphology.
}
{We found the discs to be formed by conserving the specific angular momentum in accordance with the classical disc formation model. The specific angular momentum of the stellar central spheroid correlates with the dark matter halo angular momentum and determines a power law.
Our results suggest that gas accretion to rebuild the disc component is a necessary but not  sufficient ingredient for a disc-dominated galaxy in a 
hierarchical clustering scenario. The survival of the discs is tightly linked to their subsequent history of accretion, as previously reported.
Minor mergers play an important role. It is more likely that they will destroy the disc and feed the central spheroid if their angular momentum is misaligned with respect to that of the  main galaxy. Conversely, if they are aligned, they can contribute to the disc formation. 
}{} 


\titlerunning{The morphology of galaxies}
\authorrunning{Pedrosa et al.}

\maketitle

\section{Introduction}

The morphology of galaxies is the result of a complex combination of  physical mechanisms acting at different scales and 
with different efficiencies.
Observations show well-defined correlations between morphology and
astrophysical properties such as star formation efficiency, metallicities, and sizes.
Understanding these observations within a  galaxy formation model is a challenging task.
In the current cosmological paradigm, $\Lambda$-cold dark matter ($\Lambda$-CDM),  galaxy formation follows
 highly non-linear paths, involving
 the interplay of different physical processes, which finally determine the morphology of galaxies.
 Numerical simulations are a powerful tool to tackle such a complex problem.
In the last decades, the  development of more sophisticated sub-grid  models for baryonic physics have contributed to simulate more realistic galaxies.
 Improved supernova (SN) feedback schemes \citep{okamoto2005,scan2008,Agertz2011} together with a deeper understanding of the regulation of the star formation \citep{Guedes2011,Agertz2011}, and the action of SN outflows  \citep{Brook2011, Governato2012, stinson2012} generate galaxies that better agree with observations.
High numerical resolution simulations show that gas might flow into galaxies along filaments, feeding the galaxies
directly with high angular momentum gas \citep{keres2005, dekel2009}.
However, even within this encouraging scenario it is still intensely debated how galaxies formed in dark matter haloes of different masses and which physical parameters are crucial in determining their final morphologies.
 
In the standard disc formation scenario \citep{fall1980,MMW98}, the gas cools and flows toward the centre of the dark matter halo, conserving the specific angular momentum and forming rotational supported structures.
Mergers have been shown to be able to easily destroy the discs and form spheroidal structures,
 principally during major mergers. First proposed by \citet{Toomre1972}, mergers have been  considered as one of the main mechanisms responsible for the morphological evolution of the galaxy.  However, major mergers have been shown to be rare in $\Lambda$CDM \citep[e.g.][]{deRossi2012}, except for  massive galaxies.
Disc instabilities and minor collisions have also been claimed to be efficient processes to drive material from the disc
towards the central regions \citep[e.g.][]{bh96,dominguez98,tissera2000,Mihos2004,scan2009}. Minor mergers have also been claimed to
play an important role in the formation of elliptical galaxies at $z <2$ \citep[e.g.][]{khochfar2006,naab2009,hopkins2010}.
Recently, observations and simulations suggest that the formation of stellar clumps in gas-rich unstable discs
might be another viable channel for classical 
bulge formation \citep[e.g.][and references therein]{perez2013, bournaud2013}.

Analysis of simulated galaxies with a variety of morphology suggests that alignment of the angular momentum of the accreted material with the  angular momentum of the disc
plays an important role  in determining galaxy morphology \citep[e.g.][]{scan2009}. Recently, \citet{sales2012} reported  the formation of disc-dominated galaxies to be correlated with  coherent alignment of the angular momentum of baryons that are accreted over time. \citet{martig2012} found that galaxies with the largest discs experienced no major mergers since $z \sim  2$ and had low and constant gas accretion histories, which keeps a stable angular momentum direction. These authors obtained disc-to-total stellar mass ratios ($D/T$) ranging from $~ 0.4$ to $~0.8$ for  simulated galaxies in a cosmological simulation run with the RAMSES code \citep{Teyssier2002}. \citet{Agertz2011} obtained Milky Way-like spirals with similar $D/T$ values.
\citet{Guedes2011}  obtained a $D/T$ fraction of 0.7 for Eris simulation, a very high-resolution simulation of a Milky Way-size galaxy \citep{Diemand2008}.

In this work, we aim at examining in more detail the evolution of the angular momentum content of galaxies with
different morphologies at $z=0$, but with no recent major merger (1:3).
It is well-known that major mergers have violent effects on the dynamics of discs and strongly affect them. Hence, the analysis of  a sample of galaxies with no recent major merger but different final morphologies might contribute to obtain deeper insight into the role played in the angular momentum evolution by other physical processes 
such as  minor mergers and the orientation of the accreted material.

Motivated by this reasoning, we analysed 15 galaxies that have  quiescent recent history of evolution in the sense
that none of them has experienced a major merger since $z \sim 3$.  Our aim is to study the differences in their
histories of formation  and  to understand the origin of their diverse morphology.
For this purpose, we used a simulation from the Fenix Project to analyse the evolution of a galaxies with different morphologies. The analysed simulation represents a typical field region. The Fenix project aims at studying the chemical evolution of galaxies (Tissera et al., in preparation). 
Our sample is composed of 15 galaxies resolved with more than  20000 particles within the galaxy radius.
These galaxies exhibit a variety of morphologies, with
stellar mass $D/T$ ratios varying from $ \sim 0.3$ to $  0.7$. These ratios are within the range found by previous works \citep{Teyssier2002,Diemand2008,Agertz2011}. 
Nevertheless, there is a lack of systems with  $D/T$ ratios close to unity (i.e. disc-dominated ones), which suggest an excess of stars formed in early stages of galaxy evolution \citep[see discussion by][regarding the star formation activity in this simulation]{DeRossi2013}.
We study the angular momentum content, morphology evolution and the mergers  events of the 15 chosen galaxies 
and their progenitor systems in detail. 

This paper is organized as follows: in Section \ref{sec:simus} we describe the numerical simulations. In Section \ref{sec:main} the main properties of the  galaxy sample and their angular momentum content for the central spheroids, disc, and total baryonic masses are discussed. In Section \ref{sec:formation}, we analyse the evolution of morphology in relation to the properties of the accreted systems: angular momentum orientation and relative masses.  In Section \ref{sec:conclusions} we summarize our main results.

\section{Numerical experiments}
\label{sec:simus}

We analysed a cosmological simulation from the set of the Fenix project (S230A, Tissera et al. in preparation).
This simulation has been also used by \citet{deRossi2012} and \citet{deRossi2010} to  study the rotation curves and the Tully-Fisher relation of galaxies, and by \citet{DeRossi2013} to  analyse the regulation of the star formation in haloes of different masses.
The initial condition represents a  typical field region of the $\Lambda$-CDM Universe with $\Omega_{\Lambda}=0.7$, $\Omega_{\rm m}=0.3$, $\Omega_{b}=0.04$, a normalization of the power spectrum of $\sigma_{8}=0.9$ and $H_{0}= 100 h \ {\rm km} \ {\rm s}^{-1}\ {\rm Mpc}^{-1}$, with $h=0.7$. 
A comoving cubic box of $14$ Mpc described with $2 \times 230^3$ initial particles, achieving a mass resolution of $5.9\times 10^{6}{\rm h^{-1}\ M_{\odot}}$ and $9.1\times 10^{5}{\rm  h^{-1}\  M_{\odot}}$ for the dark matter and initial gas particles, respectively. The adopted gravitational softening length   was 1.24 kpc~h$^{-1}$. The minimum gas smoothing length was 0.10, in units of the gravitational softening length.  We used  $\sim 33$ neighbouring particles for the hydrodynamics estimations.

The simulations were performed by using the code {\small GADGET-3}, an update of {\small GADGET-2 } \citep{springel2003, springel2005}, optimized for massive parallel simulations of highly inhomogeneous systems. This version of {\small GADGET-3} includes treatments for metal-dependent radiative cooling, stochastic star formation (SF), chemical enrichment, and the multiphase model for the interstellar medium (ISM) and the supernova (SN) feedback scheme of \citet{scan2005,scan2006}. 
This SN feedback model is able to successfully trigger galactic mass-loaded winds without introducing mass-scale parameters, which makes it especially suitable for the study of galaxy formation in a cosmological context. 

The multiphase model for the ISM allows the co-existence of diffuse and dense gas phases, as explained by \citet{scan2006}. In this model, each gas particle defines its cold and hot phases by applying local entropy criteria, which  allows gas particles to decouple hydrodynamically from particularly low-entropy particles, if they are not part of a shock front. Each cold-gas particle stores the energy in a reservoir until they fulfil the conditions to be promoted and to join the hot phase. Our SN feedback scheme does not include free parameters that depend on the global properties of the given galaxy (such as the total mass and size), which makes it suitable for cosmological simulations where systems with different masses form in a complex way.

The chemical model describes the enrichment by Type II (SNII) and Type Ia (SNIa) supernovae according to  the nucleosynthesis productions  of \citet{WW95} and \citet{thiel1993}. 
All stars more massive than 8 M$_{\odot}$ are assumed to end their lives as SNII, while for the rate of SNIa, an observationally motivated value of 0.0015  is adopted  \citep{mosco2001}. 
To estimate the number of stars of a given mass, we assumed a Salpeter initial mass function. The time-delay for the ejection of material of SNIa is randomly selected within the range of  [0.1, 1] Gyr, while SNII are assumed to explode within the integration time step.
The energy released by each SN event ($0.7 \times 10^{51 } $erg s$^{-1}$) is distributed into the cold and hot gas media surrounding a star particle.   A fraction $\epsilon_{c}$ of the energy released by each SN event is injected into the cold gaseous medium, while the rest is pumped into the hot phase.  
For S230A $\epsilon_{c} =0.5$, chemical elements are distributed using the same $\epsilon_{c}$, but are injected into the cold and hot phase as soon as they are synthesized.

\section{Main  galaxies}
\label{sec:main}

We study in detail  the main galaxies in virialized haloes identified by  using a friends-of-friend and the {\small SUBFIND} algorithm \citep{springel2001}.  A main galaxy is defined as the more massive system within a virial potential well.
The physical properties for  the main galaxies are calculated within the  galaxy radius ($r_{\rm gal}$) defined as  the one that enclosed $\sim 80$ per cent of the baryonic mass of a galaxy. 
Simulated galaxies with more than 2000 total particles within $r_{\rm gal}$ in S230A have also been analysed by \citet{deRossi2010}, who found  them to reproduce the observed baryonic Tully-Fisher relation.
At $z=0$, 
the star formation efficiencies reproduce observed motivated trends \citep{moster2013}, as reported by \citet{DeRossi2013}.
 However, as pointed out by these authors, these simulated galaxies  still show an excess of stars, which can be tracked to a highly efficient  transformation of gas into stars at high
redshift.  This is an open problem, although there are results that suggest that the early action of stellar feedback from massive
stars \citep{stinson2012,aumer2013} or even high-energy sources such as 
X-ray high-mass binaries \citep{power2013, artale2013} might help to regulate the star formation activity at very high redshift.
Even with this caveat in mind, it is possible to learn about galaxy formation by analysing the differences in the final morphologies in relation to the formation history of the galaxies \citep[e.g.][]{scan2009,martig2012,sales2012}.

 To reduce numerical resolution problems, we restricted our analysis to the main galaxies resolved with more than 20000 baryons within $r_{\rm gal}$. This criterion yielded a subsample of 15 main galaxies (hereafter, the 15 chosen galaxies),
with virial masses ranging from $\sim 3\times 10^{11} M_{\odot}$ to $\sim  3 \times 10^{12} M_{\odot}$, all of them resolved with $\sim 100000$ to $\sim 900000$  total particles.
For a fully cosmological simulation, these haloes are quite well resolved and allow us to analyse the properties of the galaxies \citep[see also ][]{arepo2014}. 
We also know from previous analysis of  a higher resolution simulation, 
with a factor of eight in particle number and the same physical model, that  the properties of galaxies are well represented, as discussed in detail  by \citet{deRossi2012,DeRossi2013}.

By adopting the criteria proposed by \citet{tissera2012}, we  decomposed the stellar populations into three dynamical components: central spheroid, disc, and stellar halo. For this purpose, we estimated the parameter $\epsilon$ of the  star (gas) particles defined as $\epsilon = j_{z} /j_{z,max}(E)$, where $j_{z}$ is the angular momentum component in the direction of the total angular momentum, and $j_{z,max}(E)$ is the maximum $J_{z}$ over all particles of given binding energy, $E$. 
A particle on a perfect prograde circular orbit in the disc plane has $\epsilon=1$. 
We considered particles with $\epsilon$ higher than 0.5 to be part of a disc. 
This is a low threshold, but it is consistent with the gravitational length used to run these simulations ($\sim 1.2$ kpc~h$^{-1}$). 
 The particles  that did not satisfy this requirement were taken to belong to the spheroidal component.
 By inspecting the $\epsilon - E$ plane, we checked wether the adopted limits were suitable to individualize the rotationally supported components. We also distinguished between the central spheroid and the stellar  halo components.
 The central spheroid is composed of the the most  bounded particles, while the less bounded ones are considered to form the stellar halo. Following \citet{tissera2012}, we adopted  the minimun $E$  value of the particles that fill a casket at half  of $r_{\rm gal}$ to distinguish the central spheroid from the stellar halo. Particles less bounded than these were considered to be part of the halo. This energy-based criterium was successfully used with the higher resolution simulations of the Aquarius galaxies \citep{tissera2012,tissera2013}. Nevertheless, note that  the stellar haloes do not contribute largely to the mass in the central parts that are dominated by the most bounded particles. Hence,  appreciable variations of these criteria do not  affect our
estimations.

\begin{figure}
\hspace*{0.1cm}\resizebox{4cm}{!}{\includegraphics{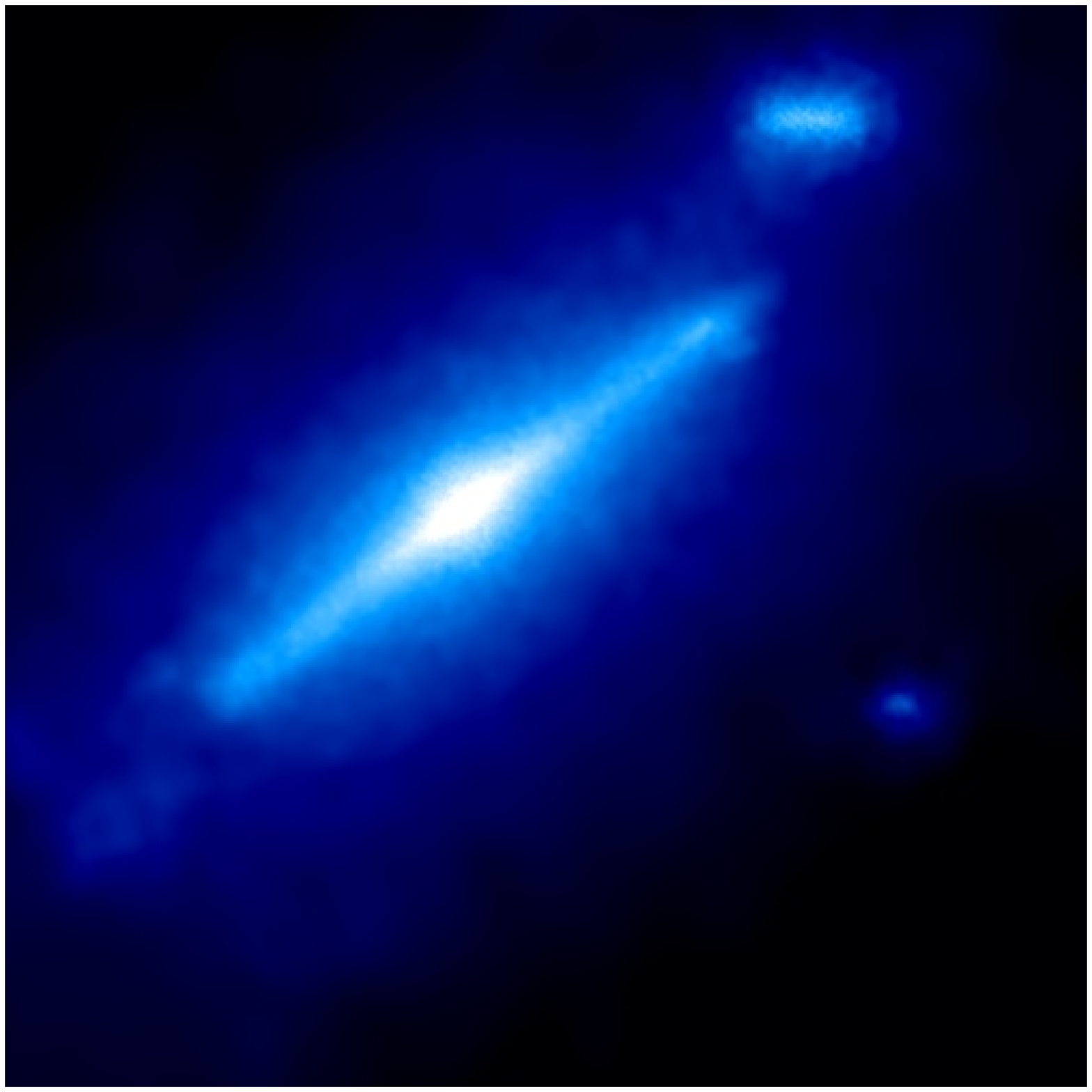}}
\hspace*{0.1cm}\resizebox{4cm}{!}{\includegraphics{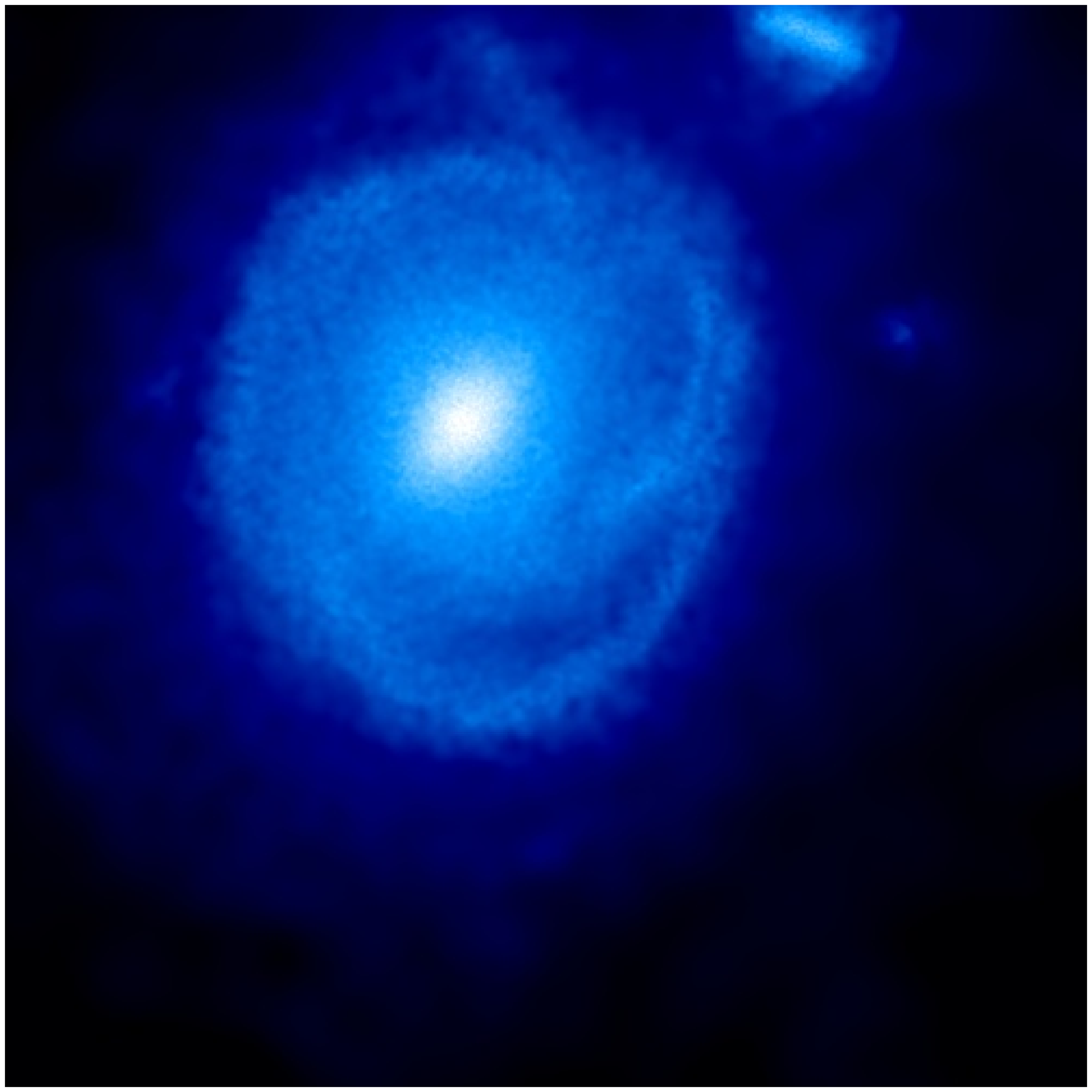}}\\
\hspace*{1.1cm}\resizebox{6cm}{!}{\includegraphics{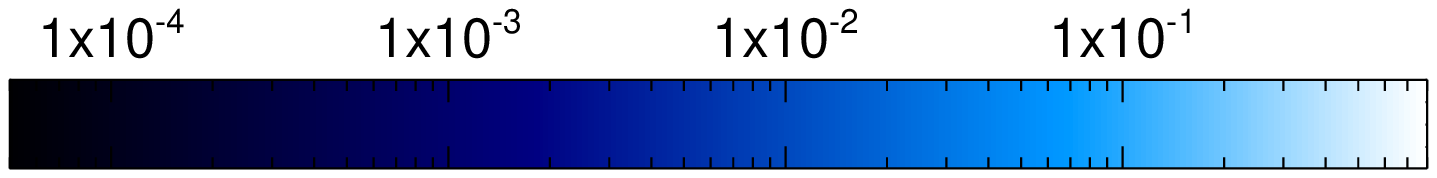}}\\
\hspace*{0.7cm}\resizebox{6cm}{!}{\includegraphics{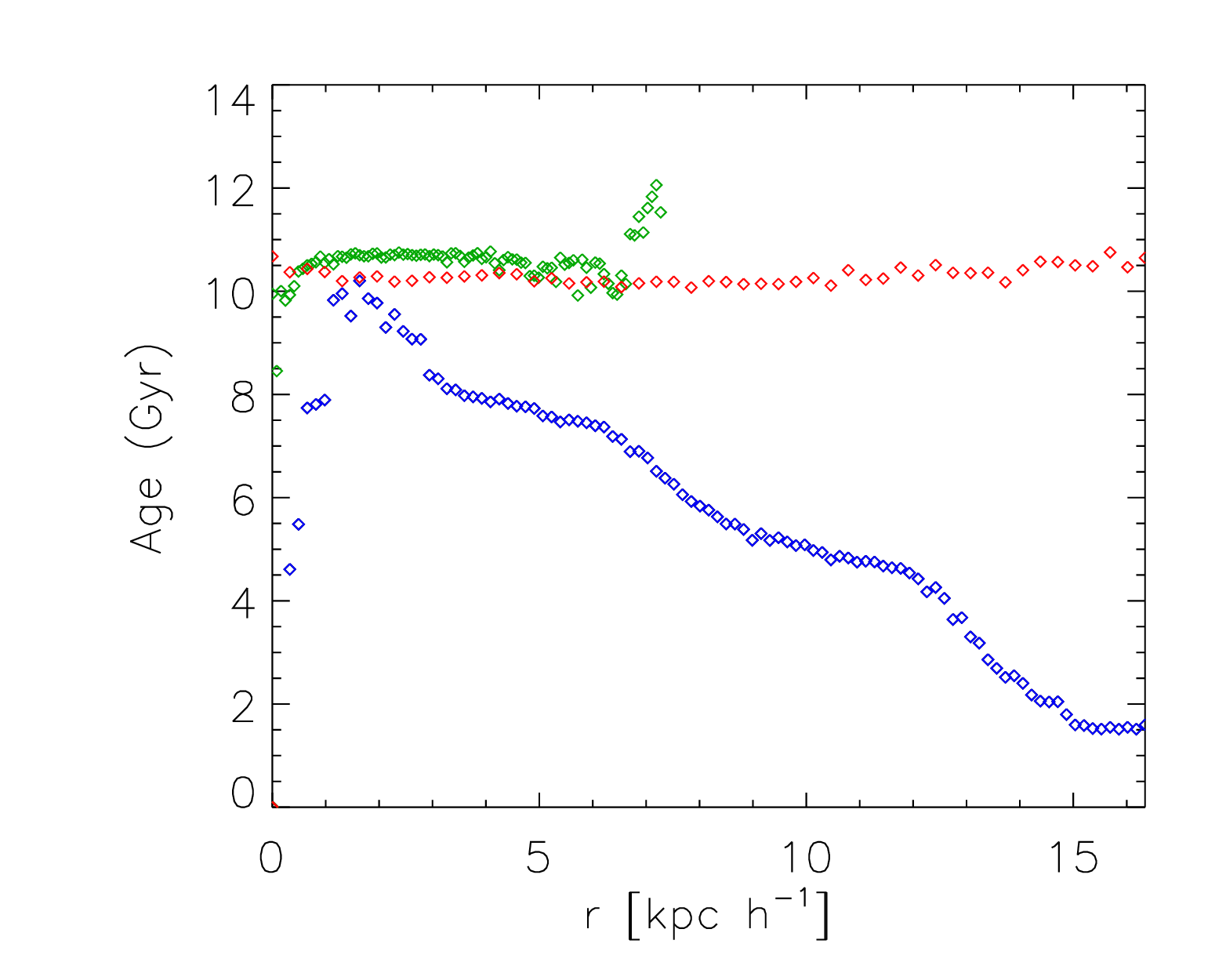}}
\hspace*{0.7cm}\resizebox{6cm}{!}{\includegraphics{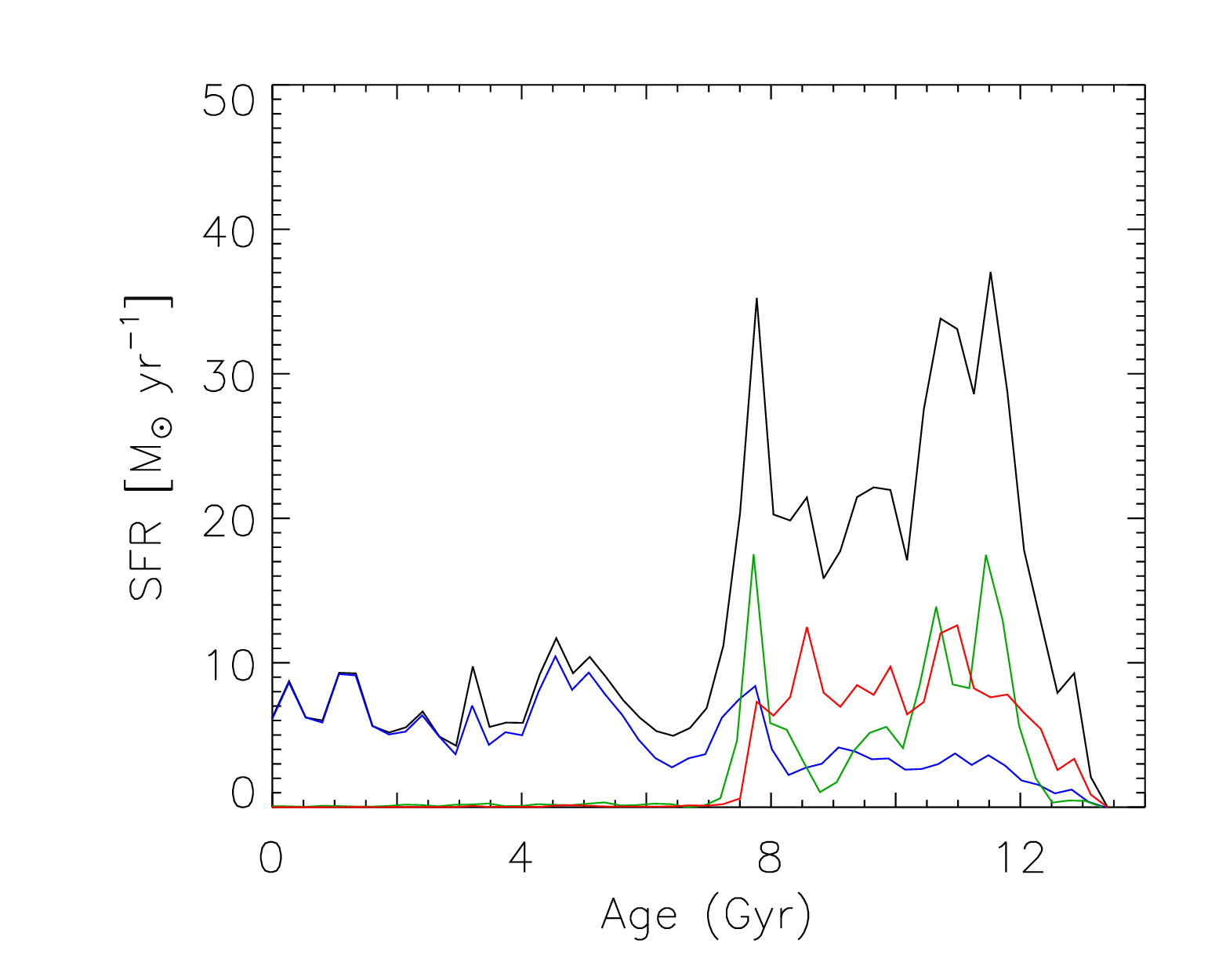}}
\caption{Edge-on  and face-on density maps for a typical galaxy of our catalogue (G1 in Table \ref{tab:tab1}) in simulation S230A (upper panels). 
 The color bar shows stellar density in units of $10^{10}$ M$_\odot $  kpc$^{-3}$. The middle and lower panels show the age gradients  and star formation rate  for stars in the central spheroid (green), the disc (blue), the stellar halo (red). The black solid line in the lower panel correspond to the total star formation rate. }
\label{fig:mapas_den}
\end{figure}

In Fig. \ref{fig:mapas_den} we show the face-on and edge-on projected stellar density maps for galaxy G1 in Table \ref{tab:tab1}, which has a large stellar disc at $z=0$.
 In the lower panel we show the age profiles for the different stellar components as a function of radius.
 It can be appreciated that stars in the disc component determine a well-defined negative age gradient consistent 
with an inside-out  history of  formation. This is  a common feature of  our simulated galaxies that were able to retain large stellar discs. Conversely, the central spheroid and halo stellar components do not show a clear age profile and are populated by  old stars with mean ages of $\approx 10$ Gyr. Stars in the central spheroids are systematically older than those in the stellar haloes.
 This reflects their different assembly history: the central spheroids are mainly formed in situ, within the  deepest potential wells, while the stellar haloes are principally assembled from small substructures \citep{tissera2012}.  Even though the SFRs of  the central spheroid and stellar halo  
show the highest activity at very early times, the former transformed most of their stars in a shorter burst. For the system shown in  Fig. \ref{fig:mapas_den}, the stars in the central spheroid were mainly formed 
 between  $\sim 10$ and  $\sim 12$  Gyr ago  and  with  a rejuvenating  short-lived peak at $ \sim 8$ Gyr. The stellar halo exhibits a more extended star formation history from $\sim 12$ to $\sim 8$ Gyr ago.  
As expected, the disc component experiences the  higher star formation activity at low and intermediate redshift with some contribution from old stars. For this system,  the mean age of the stellar populations in the  disc is $\sim 5$ Gyr and the contribution  of stars with ages older than  $8$ Gyr   represents  $\sim 23 $ per cent of the total stellar mass. 
The inside-out disc formation scenario for normal disc galaxies has been confirmed by several N-body/SPH simulations \citep[e.g.][]{Guedes2011, Agertz2011}.
The standard disc formation scenario \citep[see e.g.][]{MMW98} provides a physical explanation for observational results where negative-age gradients are commonly found. In particular, recent detailed observations
of the {\small CALIFA} survey reported disc components to have negative-age gradients and  bulges to be consistent with older populations with flatter profiles in spiral galaxies with  total stellar masses larger than  $10^{11} $M$_{\odot}$ \citep[][see also \citet{macarthur2004}]{califa2013}.

 We performed  a double component decomposition for each of the 15 chosen galaxies. For the discs, we adopted an exponential profile 
with a scale-length $R_{\rm d}$. For the central spheroid, we assumed a Sersic law \citep{sersic1963} with an index $n$  and a scale-length $R_{\rm b}$.  As can be seen in Table \ref{tab:tab1}, we found a S\'ersic index   lower than $ n \sim 2$, consistent with pseudo-bulges.  Numerical results reported by \citet{martig2012} were similar and seem to agree with observational findings of local disc galaxies where mean $n$ values lower than two are reported \citep{laurikainen2007,weinzirl2009}.
 In  Table \ref{tab:tab1}, we also include the  $D/T$ fraction  defined as $M_{\rm disc}/(M_{\rm disc}+M_{\rm spheroid})$, where $M_{\rm disc}$ and $M_{\rm spheroid}$ are the dynamical masses of the disc and central spheroidal components obtained by the kinematic decomposition described above. 

\begin{figure}
\resizebox{8cm}{!}{\includegraphics{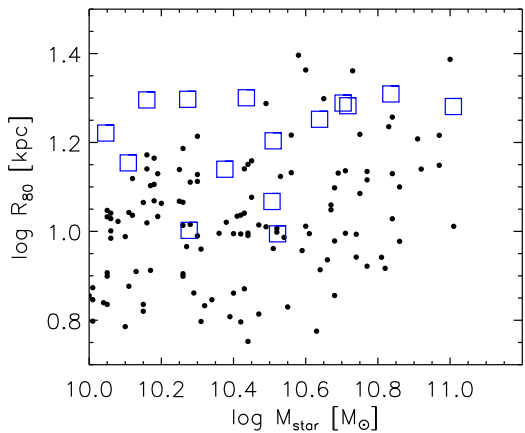}}
\hspace{-0.2cm}
\caption{ Optical radius as a function of the stellar mass of the simulated galaxies (blue open squares) and the observational
estimates of \citet[][black circles]{reyes2011}.  } 
\label{fig:reyes}
\end{figure}

We used the recent observational results from \citet{reyes2011, reyes2012} to compare the size-stellar mass relation determined by our simulated disc galaxies. Because
the authors estimated the optical radius $R_{80}$ for their observed galaxies, we defined  $r_{80}$  as the radius that encloses $\sim 80$ of the stellar mass of a given galaxy. In doing so, we asumed a constant mass-to-light ratio \citep{tissera2000} In Fig. \ref{fig:reyes}, we plot $r_{80}$ as a function of the galaxy stellar mass for the simulated galaxies and the observed values reported by \citet{reyes2011}. The simulated galaxy sizes generally agree with these observations,  although they tend to populate the upper part of the observed distribution because of the action of the strong SN feedback adopted in this run, which produces more extended baryon distributions \footnote{\citet{scan2006} discussed in detail the effects of the SN feedback model we use in S230  on baryons by running the same halo with different SN feedback parameters. In addition, in \citet{deRossi2010} we compared S230 with a run where the SN feedback was switched off. These analyses show that only when SN feedback was on, extended 
discs were produced.}. Unfortunately, the  high numerical resolution limit imposed on the selected galaxies left us with a low number sample to assess the existence of a correlation between galaxy scale and galaxy mass. This price is worth paying because we aim at analysing the evolution of the angular momentum across time, which can be easily affected by a low number of particles.

\begin{table*}
  \begin{center}
  \caption{Main characteristics of our subsample of the 15 most massive objects in S230A: the number of dark matter ($N_{\rm dark} $), star ($N_{\rm star} $), and gas ($N_{\rm gas}$) particles within the virial radius, the  virial radius ($R_{\rm 200}$), the virial mass  ($M_{\rm 200}  $),  the stellar ($M_{\rm star}$) and baryonic ($M_{\rm bar}$) masses within the baryonic radius, the Sersic index for the central spheroid ($n$), the  disc scale length ($R_{\rm d}$), and the disc-to-total-mass ratio ($D/T$).}
  \label{tab:tab1}
 {
  \begin{tabular}{|l|c|c|c|c|c|c|c|c|c|c|}\hline
Galaxy  &  {$N_{\rm dark} $} & {$N_{\rm star} $} & {$N_{\rm gas}$} & {$R_{\rm 200}$}  & {$M_{\rm 200}  $} & { $M_{\rm star}  $}  & { $M_{\rm bar}  $}  & {$n$}  & { $R_{\rm d}$}  &  $D/T$\\
  & & & &  (kpc  $h^{-1}$) &  ($10^{11} M_{\odot} h^{-1}$) & ({$10^{10} M_{\odot} h^{-1}$})  & ({$ 10^{10} M_{\odot} h^{-1}$})  &  &  (kpc  $h^{-1}$) &\\ \hline
G1 &  437013   &   255241   &   192495  &     231.1  &   25.9 & 7.7  &   8.0  &     1.1    & 4.1 &0.56\\
G2 &  156183   &    92670   &    28190  &     162.2  &   9.3 & 3.1  &   3.3  &     2.2    & 2.5 &0.55\\
G3 &  228052   &   120262   &    82041  &     185.2  &   13.5  & 4.0  &   4.3  &     1.5    & 3.2 &0.46\\
G4 &  162563   &    72102   &    44930  &     164.6  &   9.6  & 2.3  &   2.4  &     1.4    & 2.1 &  0.71\\
G5 &   71815   &    48399   &    14528  &     125.5  &   4.3  & 1.8  &   1.8   &     1.3    & 1.5 & 0.34\\
G6 &  181662   &   113855   &    52922  &     171.6  &   10.8  & 3.9  &   4.3  &     1.9    & 3.1 & 0.56\\
G7 &  117001   &    60969   &    49564  &     148.6  &   6.9  & 2.1  &   2.4  &     1.3    & 5.1 &0.44\\
G8 &   82096   &    40720   &    23221   &     131.1  &   4.9  & 1.4  &   1.7  &     1.3    & 5.6 &  0.39\\
G9 &   69753   &    31088   &    15567   &     123.8  &   4.1  & 1.1  &   1.5  &     1.3    & 4.0 & 0.52\\
G10 &  260771   &   154477   &    74724  &     193.2  &   15.5  & 5.0  &   5.3  &     1.3    & 4.2 & 0.47\\
G11 &  136695   &    71156   &    28907  &     155.1  &   8.1  & 2.4  &   2.5  &     1.4    & 1.9 & 0.59\\
G12 &  117048   &    49347   &    27641  &     147.1  &   6.9  & 1.4  &   1.5  &     1.1    & 2.8 &  0.50\\
G13 &  129747   &    71671   &    36577  &     153.0  &   7.7 & 2.6  &   2.7  &     1.5    & 1.7 & 0.66\\
G14 &  62344   &   27559    &   14604   &     119.2  &   3.7  & 0.9  &   1.0  &     1.3    & 3.5 &  0.49\\
G15 &  49406   &   24659   &  7370     &     110.0  &   2.9 & 0.8  &   0.9  &     1.3    & 1.4 & 0.27\\ \hline
  \end{tabular}
  }
 \end{center}
\vspace{1mm}
\end{table*}

\subsection{Angular momentum content}
\label{sec:angmom}

The angular content of the galaxies in relation to that of their host dark matter haloes has been analysed in previous works, as discussed in the Introduction. It has been  shown that the disc components tend to form by conserving the specific angular momentum as postulated by the standard disc formation model. In this work,  we extend this analysis by examining the angular momentum content of the disc as well as of the central spheroidal components  in galaxies with different $D/T$ ratios.  

We estimate the specific angular momentum of the stellar and gaseous disc components within $r_{\rm gal}$ and  of the dark matter haloes within $R_{200}$.
 In the model of \citet{MMW98}, it is assumed that the mass that settles into the disc is a fixed fraction of the dark matter halo mass,  $m_{\rm d}=M_{\rm D} /M_{\rm H}$. Similarly, its angular momentum is assumed to be  a fraction $j_{\rm d}$ of that of the halo defined as $j_{\rm d}=J_{\rm D} /J_{\rm H}$. Following these concepts, we calculated  $m_{\rm d}$ and $j_{\rm d}$ for the stellar and gaseous discs. We also applied these definition for the total baryonic mass within  $r_{\rm gal}$ .

\begin{figure}
\resizebox{8cm}{!}{\includegraphics{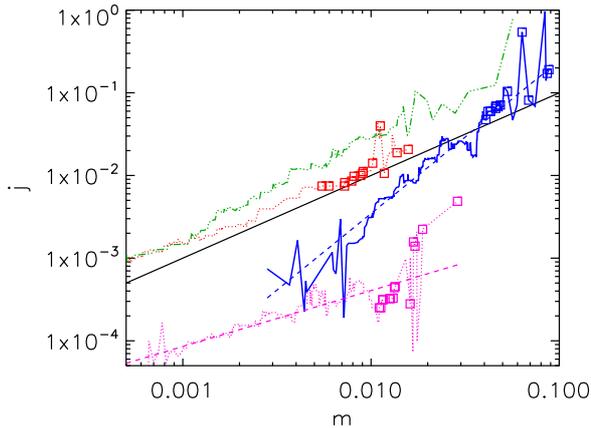}}
\hspace{-0.2cm}
\caption{Median angular momentum fraction ($J_{\rm D} /J_{\rm H} = j_{\rm d}$) for  the stellar (red dotted line) and  gaseous (green dashed-dotted line) disc components  as a function of the ratio between  the stellar and  gaseous disc masses and of the dark matter haloes.
We also show the same relation for the stellar central spheroids  (pink dashed line),  which determines  a well-defined correlation, but still consistent with weaker angular momentum conservation.  The relation for  the total  baryonic mass is also included  (blue solid line). The blue dashed  line shows the relation
 $j \propto m^2$.  The pink dashed  line represents the linear fit to the relation defined by  stellar central spheroids.  }
\label{fig:AngMom}
\end{figure}

As can be seen  in Fig. \ref{fig:AngMom}, we found a clear correlation between the  angular momentum of both gaseous and stellar discs and that of the haloes, with the $j_{\rm d} \propto  m_{\rm D}$ relation as expected for angular momentum conservation. Under the hypothesis of angular momentum conservation, $j_{\rm d}$ should be equal to $m_{\rm d}$ \citep{fall1980}. This assumption is based on the fact that all the material in the system experiences the same external torques before separating into two distinct components. The stellar discs show the best behaviour, while the gaseous discs have gained additional angular momentum because of the action of the SN feedback, which generates more extended gas distributions. On average, the 15 chosen galaxies in S230A have stellar discs that agree with the hypothesis of angular momentum conservation.

 In Fig. \ref{fig:AngMom}, we also show $j_{\rm s}$ versus $m_{\rm s}$ estimated by using stellar  masses of the central spheroids. As can be seen from this figure,  these stars have been formed from low angular momentum material. Nevertheless,  there  is a clear correlation between the specific angular momentum of the total stellar mass of the central spheroids and their dark matter haloes so that $j_{\rm s} \propto m_{\rm s}^{0.7}$. The correlation is noisier than that determined by the discs, but is still significant ($r \sim 0.6$). The angular momentum content of the stellar components in both the central spheroids and the disc components can be affected by the mergers and interactions (e.g. if there is angular momentum transfer between the components and the halo) and by the SN feedback, which helps to regulate the star formation activity, but also can affect the dynamical evolution of the gas from which the next generation of stars will form.

 The specific angular momentum of the total baryonic component of a galaxy and that of its dark matter halo determine a relation close to $j_{\rm bar} \propto m_{\rm bar}^{2}$. This trend agrees with the results reported by \citet{sales2012},  although the SN feedback models are different. The change in the slope relation in our simulated galaxies may be due to the effects of the redistribution of angular momentum between the components during violent events \citep[e.g.][and references therein]{pedrosa2010}, combined with the action of SN feedback that transfers angular momentum to the gas component and drives powerful winds.  It is expected that these winds remove preferentially low angular momentum gas \citep[e.g.][]{Brook2011}. 
 
Our results show that if a disc component is able to form, it does so according to the classic formation scenario via conservation of the specific angular momentum, in agreement with previous works.  The material that forms the disc conserved the specific angular momentum  obtained via tidal torques  by the initial density perturbations. We also find that stellar central spheroids,  which formed from low angular momentum material, determine a clear correlation with the  angular momentum content in the dark matter halo. These results might contribute to explain the observational size-mass relation reported by  \citet{kravtsov2013}. As suggested by this author, the angular momentum content might play a critical role in setting the size of galaxies of different morphologies, even spheroidal-dominated ones.
A comprehensive analysis of the angular momentum evolution of the dynamical components and the relation with the sizes will be performed
in forthcoming paper  (Pedrosa \& Tissera, in preparation).

\section{Formation and survival of the stellar discs}
\label{sec:formation}

We quantified the evolution of the morphology  by  analysing the $D/T$ ratios as a function of redshift and the correlation of this parameter with the assembly history of the simulated galaxies. As galaxies evolve, the $D/T$ ratios of the progenitor systems can change, reflecting their different assembly histories. 
We adopted $D/T =0.5$  as a threshold  to separate our simulated galaxies into two groups:  disc and spheroidal-dominated systems. 
At $z=0$ , the 15 chosen galaxies  have $D/T$ values ranging from $\sim 0.3$ to $\sim 0.7$.

We constructed the merger tree for each simulated galaxy at $z=0 $ and followed the evolution of the main progenitor branch. At each available cosmic time, we defined the main progenitor as that which exhibits the most significant baryonic component. All other smaller systems in the merger tree were considered satellites that will eventually merge onto the main branch \citep[see also][]{deRossi2012}.

\begin{figure}
\resizebox{8cm}{!}{\includegraphics{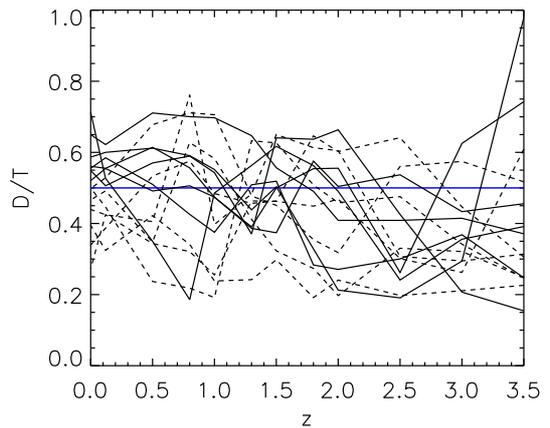}}
\hspace*{-0.2cm}
\caption{$D/T$ ratios for the progenitor galaxies of the 15 chosen galaxies as a function of the redshift. Those with $D/T_{z=0} > 0.5$ are shown by continuous lines and those with  $D/T_{z=0} < 0.5$  by dashed lines.} 
\label{fig:DBevol}
\end{figure}

To examine the evolution of the $D/T$ ratios, we  performed  the kinematical decomposition explained in Section 3 to the progenitor systems  of each selected galaxy up to $z \sim 3.5$. For the purpose of illustration, we show their evolution with redshift in Fig. \ref{fig:DBevol}. It is clear  that the  $D/T$  is a very dynamic parameter, which indicates that the disc can be destroyed and  re-built several times along its evolution.
Our goal is to try to underpin the  physical conditions or processes that determine these behaviours in relation to the final  morphologies.

As discussed in the Introduction, none of the 15 chosen galaxies has experienced a major merger (1:3) since $z \sim 3$, so they might be  considered to 
have  quite recent assembly histories. Nevertheless, they all end with different $D/T$ ratios. 
Hence, the differences in the final morphology have to respond to their different assembly histories.

\begin{figure*}
\resizebox{5.5cm}{!}{\includegraphics{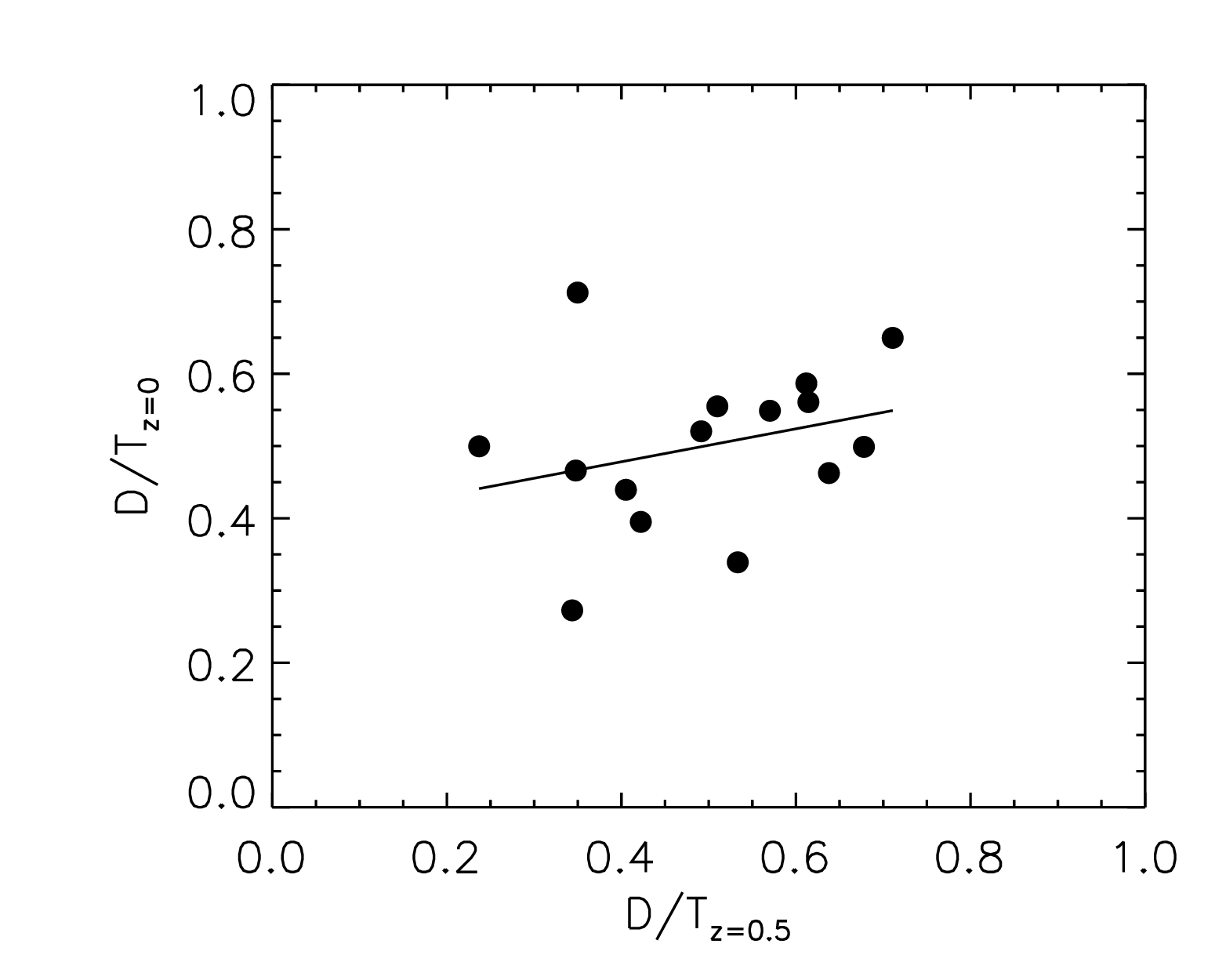}}
\resizebox{5.5cm}{!}{\includegraphics{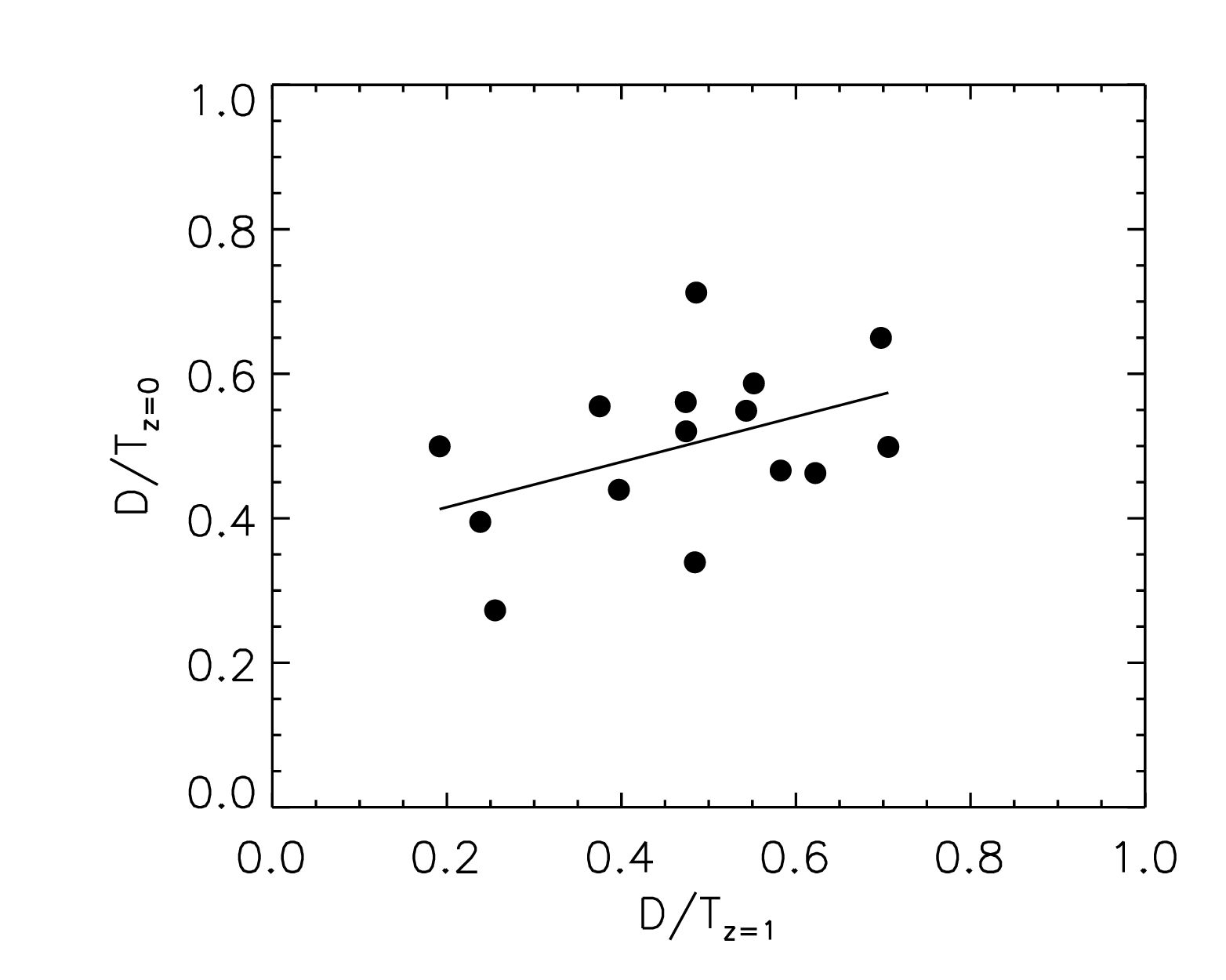}}
\resizebox{5.5cm}{!}{\includegraphics{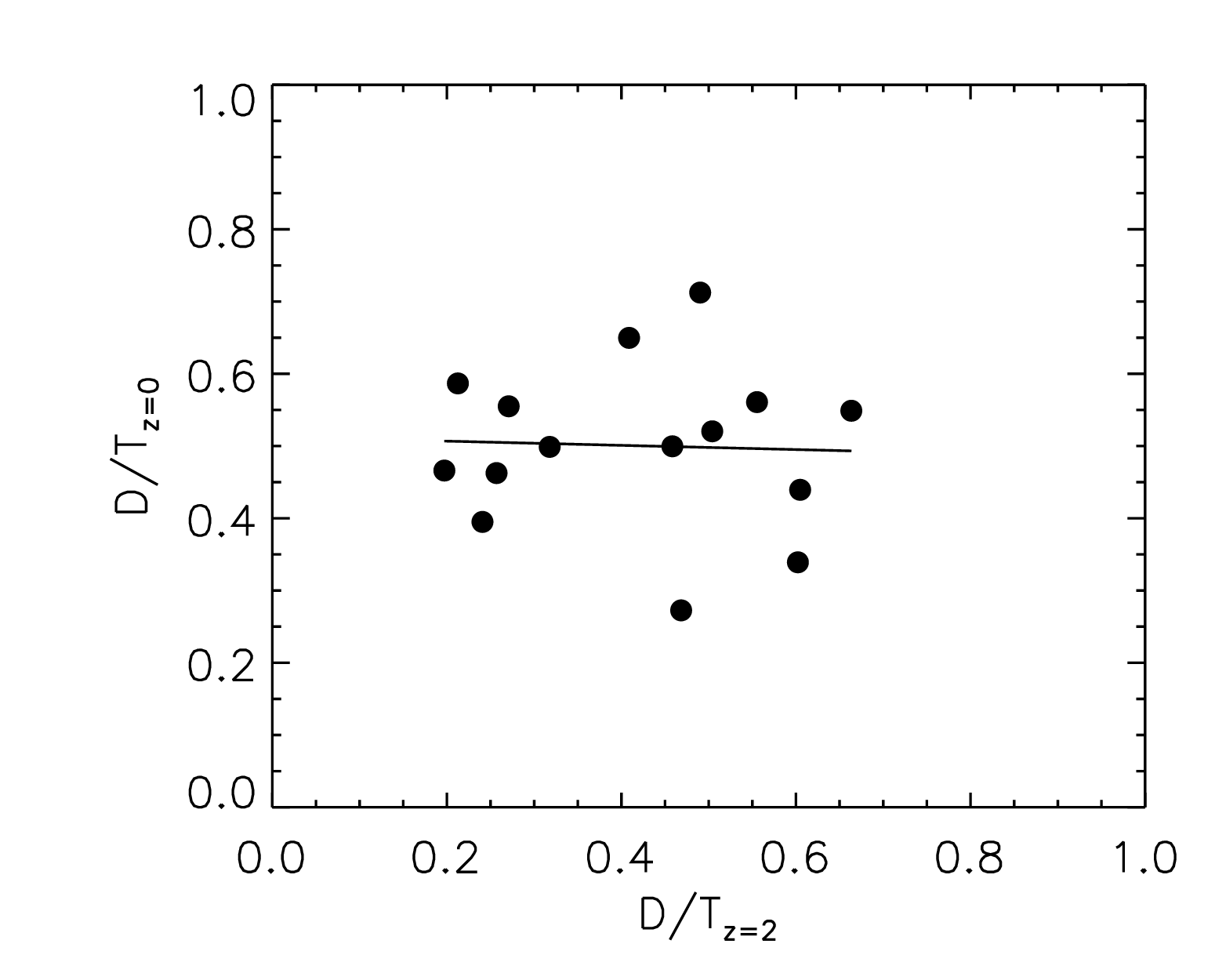}}
\hspace*{-0.2cm}
\caption{Correlation between the $D/T$ ratios  of the simulated massive  galaxies in S230A at $z=0$ and those of their progenitor systems at $z=0.5, 1$ and  $z=2$. 
The solid lines are regression fits to  the data. } 
\label{fig:DBzDB0}
\end{figure*}

We explored the correlation between the $D/T_{z=0} $ and those of  the progenitor systems measured at $z=0.5$, $z=1$ and $z=2$.
As shown  in Fig. \ref{fig:DBzDB0}, there is a global trend for the simulated galaxies to have their morphology in place since $z \sim 1$. However, by  $z \sim 2$ the  correlation is lost, suggesting that, at least in this simulation, the morphology of the final galaxies is 
principally determined  by  their history at $z > 1 $, and for lower redshift,  the morphologies of these simulated disc galaxies are not strongly modified, on average. We note that on an individual basis, there might be different behaviours, as can be seen from the large dispersion shown by the relations.
The correlation factors are $0.29$, $0.43$, and $-0.1$ for relations measured at  $z=0.5$, $z=1$, and $z=2$, respectively.
 On average, the evolution since $z \sim 1$ does not strongly affect the morphology of the simulated galaxies.
By using a different numerical approach, \citet{martig2012} found no correlation between the morphology at $z=0$ and at $z=1$.
  This difference
could be due to the different numerical implementations or to the  different set of analysed galaxies since in our work.

A crucial ingredient for disc formation is the availability of accreted gas needed to rebuild the disc, as shown in previous works \citep{sales2012}. We estimated the growth of $M_{\rm disc}$ and $M_{\rm spheroid}$ from $z\sim 3.5 $ to $z=0$. We found that galaxies with $D/T_{z=0} > 0.5$ have mean $M_{\rm disc}$ larger or similar to  the mean $M_{\rm spheroid}$ since $z \sim 2$, while those with $D/T_{z=0} < 0.5$ have central spheroids more massive than the discs in the same redshift range.  
For galaxies with  $D/T_{z=0} > 0.5$, we estimated that the disc components  have grown by a factor $\sim 2.5$  since $z\sim 1$  while the discs in spheroid-dominated galaxies have done so only by $\sim 1.2$. The stellar masses in the central spheroids have grown by a similar factor ($\sim 1.5$) for all galaxies, implying that the disc components have grown more in disc-dominated galaxies, as expected.
The relative increase of the disc-to-central-spheroid mass is due to the gas availability and the fact that the discs are able to survive and grow.
For spheroidal-dominated galaxies, the growth of the central spheroid is larger than that the of the disc. 
In next section, we analyse how minor  merger events can be associated to the decrease of the  $D/T$ ratio.

The estimation of the redshift at which $50$ per cent of the final stellar mass in the  discs was at place yielded that all discs in our sample have built up half their stellar masses  at $z < 2$. We detected a weak correlation between these parameters, suggesting that systems that formed their discs earlier have a higher $D/T$ ratio at $z=0$.

\subsection{Galaxy morphology evolution}
\label{sec:evolution}

\begin{figure}
\resizebox{8cm}{!}{\includegraphics{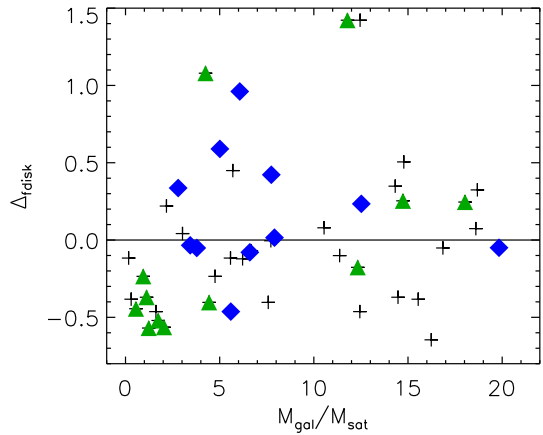}}
\caption{$\Delta _{\rm fdisc}$  as a function of the $M_{\rm gal}/M_{\rm sat}$  for the colliding systems. We show the  aligned (blue diamonds) and misaligned (green triangles) encounters. Aligned encounters are systematically associated to positive or negligible changes of the $D/T$. Crosses: encounters without defined alignment.} 
\label{fig:fd_massat}
\end{figure}

We followed the formation history of   galaxies  and studied their angular momentum content, their growth of the discs and central spheroidal components, and their merger history. 
We estimated the stellar  masses of the progenitor galaxies ($M_{\rm gal}$)  and of accreted satellites ($M_{\rm sat}$), the angle between
the total angular momenta of progenitor galaxy ($ J_{\rm gal}$) and of accreted satellites ($J_{\rm sat}$) at each available snapshot of the simulation since $z \sim 3.5$.
As galaxies evolve across time, $D/T$ experiences strong variations that might be linked with interactions with the satellites that merge with the progenitor galaxy. 
We defined  $\Delta_{\rm fdisc}  $ as the variation of the $D/T$ associated to the main progenitor of a given galaxy between two consecutive snapshots in the simulation, where the second snapshot is the one where the satellite can no longer be identified as a separate object \footnote{Since the number of available snapshots is limited, the exact time of a merger might not be available. This temporal limitation would contribute to increase the noise in the relations. Hence, only averaged trends can provide useful information.}. 
Hence,  $\Delta_{\rm fdisc} > 0$ implies a positive growth of the stellar disc component with respect to the stellar central spheroid between the two available snapshots. 
In Fig. \ref{fig:fd_massat}, we plot  $\Delta_{\rm fdisc}$ as a function of the ratio $M_{\rm gal}/M_{\rm sat}$. As can be seen, major mergers (1:3) are all associated to negative changes of $D/T$, which indicates that the central spheroids grow with respect to the discs during these events. When less massive mergers are considered, the response is different, with positive and negative variations.

We explored the impact of the relative orientation between the angular momenta of the progenitor galaxies and the accreted satellites by assuming that the interaction is aligned when $ cos(J_{\rm gal},J_{\rm sat})$  is larger than $0.7$ and  misaligned when, $ cos(J_{\rm gal},J_{\rm sat})< -0.7$ or $-0.3 >  cos(J_{\rm gal},J_{\rm sat}) < 0.3$.
We separated these extreme behaviours to highlight the effects more clearly. The interactions that cannot be classified within these categories are shown as crosses in Fig. \ref{fig:fd_massat}. It is  clear that aligned interactions produce negligible or positive variations of  $D/T$.
The median $\Delta_{\rm fdisc}$ estimated for aligned mergers is  $\Delta _{\rm fdisc} \sim 0.015$, for the misaligned mergers this is $\Delta _{\rm fdisc}  \sim -0.44$.
We note that more than $60 \%$ of the encounters that are not aligned, but have not entered our stringent misalignment classification, are
associated with negative changes in the $D/T$ ratio.

Regarding the relative masses of the interacting systems, we also found that encounters with relative stellar masses up to  $M_{\rm gal}/M_{\rm sat} \sim 6$ can have a strong impact on the galaxy morphology if they are misaligned.
 Accretions of satellites that are aligned with the angular momentum of the progenitor galaxy have a negligible impact or even increase the disc components, regardless of $M_{\rm gal}/M_{\rm sat} $, on average.  We found misaligned minor and major mergers to be associated with a strong perturbation of the disc components. 
This result agrees with previous works \citep[e.g.][]{dominguez98, scan2009}, who showed that misaligned accretions can be correlated with the disc destruction, and with the global conclusions of \citet{sales2012}, who found a correlation with the total angular momentum content at turnaround of the accreted material. Considering the results of \citet{sales2012}, we would expect that the galaxies that have large discs at $z=0$ would have  experienced  more  aligned mergers during their assembly history.

\begin{figure}
\resizebox{8cm}{!}{\includegraphics{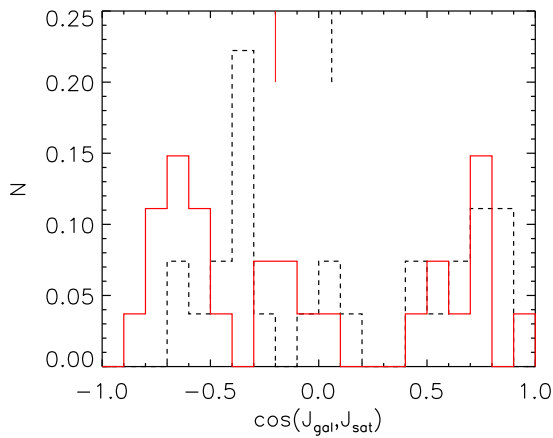}}
\caption{Histograms of the cosines between the angular momentum of the progenitor galaxies and that of the accreted satellites,  $ cos(J_{\rm gal},J_{\rm sat})$. Mergers experienced  for simulated galaxies with  $D/T_{z=0} > 0.5$ (black dashed lines) and  with $D/T_{z=0} < 0.5$ (red solid lines) are shown separately. The short lines denote the  median values for both subsamples.} 
\label{fig:histo_cos}
\end{figure}

To explore this idea, we  plot the distribution of  $ cos(J_{\rm gal},J_{\rm sat})$  for all mergers experienced by  progenitor systems of the analysed galaxies  grouped according to their final  $D/T$.  As shown in Fig. \ref{fig:histo_cos}, low $D/T$ galaxies tend to experience more misaligned encounters (i.e. negative $ cos(J_{\rm gal},J_{\rm sat})$  values). However, it should be kept in mind that a larger statistics is needed to confirm this result. The median $ cos(J_{\rm gal},J_{\rm sat})$ for galaxies with small disc components is  $-0.2$, while those with large disc components have $0.06$. Hence, simulated galaxies that were able to reach to $z=0$ with larger discs tend to  have experienced  more aligned encounters. An average positive  alignment of the in-falling structure with the progenitor during the evolutionary history  correlates with the growth of the disc, in agreement with previous results \citep{scan2009,sales2012,martig2012}.

\begin{figure}
\resizebox{8cm}{!}{\includegraphics{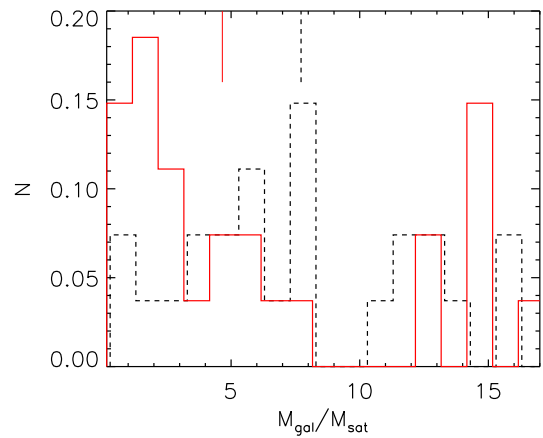}}
\caption{Histograms of the progenitor galaxy to satellite mass ratio of the 15 chosen galaxies with  $D/T_{z=0} > 0.5$ (black lines) and  with $D/T_{z=0} < 0.5$ (red). Median values for both subsamples are shown as short lines.} 
\label{fig:histo_mass}
\end{figure}

We also explored the role of the relative masses of the accreted systems,  $M_{\rm gal}/M_{\rm sat}$ because major mergers are commonly assumed to be responsible of the spheroidal formation.  We built up the histograms of  $M_{\rm gal}/M_{\rm sat} $ for the two  subsamples.
As can be seen from  Fig. \ref{fig:histo_mass}, low $D/T$ galaxies  have a higher probability to have experienced more massive mergers with a median $<M_{\rm gal}/M_{\rm sat} >\sim 5$.  For galaxies with high $D/T$ ratios, the median decreases to   $<M_{\rm gal}/M_{\rm sat} >\sim 8$.  
We have checked in both cases that the distributions are statistically different using a Kolmogorov-Smirnov test, which yielded $k_{KS}\sim 0.07$.

\begin{figure}
\resizebox{8cm}{!}{\includegraphics{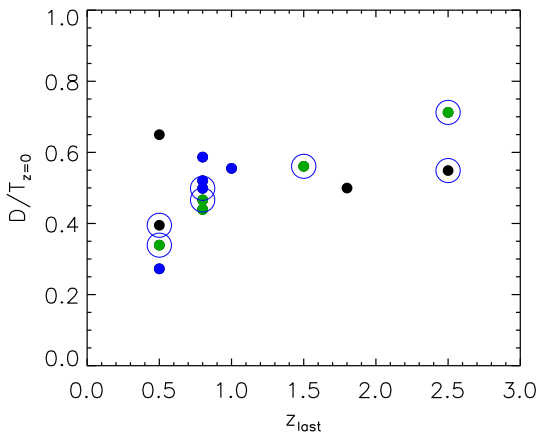}}
\hspace*{-0.2cm}
\caption{$D/T$ ratio as a function of the redshift of the last merger. Aligned (misaligned) accretion is shown as blue (green) symbols. Mergers
with relative masses larger than 1:6 are enclosed by large circles.} 
\label{fig:DB_zlast}
\end{figure}

Finally, we  estimated that the median redshift for the last merger episode is $ z_{\rm last} \sim 1.4$ for the 15 chosen galaxies with  $D/T > 0.5$, while for those with $D/T < 0.5$, the median is $ z_{\rm last} \sim 0.9$.  Hence, even though the 15 chosen galaxies have no major merger for $z < 2.5$, there is still a trend for those with larger discs at $z=0$ to experience  a quieter evolution in recent epochs. 
In Fig. \ref{fig:DB_zlast}, we plot  $D/T$ as a function of the redshift of the last merger. We included mergers with a mass ratio higher than 1:6 in circles as a reference value to distinguish  minor from very small accretions. Mergers that are aligned (misaligned) are shown in blue (green). 
Systems with destroyed discs tend to experience late mergers, which are, on average, more massive than 1:6. A  turn-fork behaviour is suggested from $z_{\rm last} \sim 1$ related to characteristics of the merger events, which needs to be  analysed with a larger sample of galaxies.

\section{Conclusions}
\label{sec:conclusions}

We analysed a sample of 15 galaxies  with virial masses between  $\sim 3 \times 10^{10}$ to  $\sim 3 \times 10^{11}$ in a $\Lambda$-CDM scenario. They all have a quiescent assembly history in the sense that none of them have experienced a major mergers (1:3) since $z \sim 2.5$. They inhabit typical field regions of the Universe. Nevertheless, they reach redshift zero exhibiting different morphologies. The analysed galaxy sample
was resolved with more than  20000 particles within the galaxy radius.
 
Our results can be summarized as follows:

\begin{itemize}
\item We found  the disc components to be formed by conserving the specific angular momentum, as expected from the classic disc formation model, so that they determine a relation  $j_{\rm d}\propto m_{\rm d}^\alpha$ with $\alpha \sim 1 $, where $j_{\rm d}$ is the ratio between the specific angular momentum of the galaxy disc and that of the dark matter halo, while $m_{\rm d}$ is the ratio for the corresponding masses \citep{MMW98}. The gaseous discs tend to show slightly steeper relation because of action of SN feedback, which determines more extended gas distributions.

We  detected that the simulated central spheroids, even though formed from low angular momentum material, determine a clear  correlation with the
 angular momentum of the dark matter haloes. For this simulation, we found $j_{\rm s}\propto m_{\rm s}^\alpha$ with $\alpha \sim 0.7 $. The correlation is noisier than
for the disc components, but it is statistically significant. The 15 chosen galaxies follow the mean trends remarkably well.  This correlation might explain recent observational results on the mass-size relation  reported
by \citet{kravtsov2013}. We postpone a deeper discussion on this point to a forthcoming paper where different SN feedback parameters will also be considered.

When all baryons were included, the relation approached $j_{\rm baryons}\propto m_{\rm baryons}^\alpha$ with $\alpha \sim 2 $, in agreement with the results of \citet{sales2012}. The change in the slope is produced by the combined effect of lost and gain of
angular momentum due to exchange between components during violent events or by the action of galactic winds \citep[see also ][]{sales2012}.

\item The analysis of the formation history of the galaxies that are dominated by the central spheroid or the disc components at $z=0$ show  a very complex path. The  alignment of the accreted satellites and the relative mass are important for the fate of the discs.
Aligned satellites are associated either with a negligible impact or with an increase of the disc mass with respect to the spheroidal mass.  
During the evolutionary path of the simulated galaxies, we detected a trend for the systems with larger disc components to have experienced, on average, more aligned  and less massive mergers than the rest. We therefore confirm results from  previous works, which used different numerical codes and SN feedback models \citep{sales2012,martig2012}. Minor mergers can also contribute significantly to the evolution of the galactic morphology depending on their relative alignment with respect to the total angular momentum of the main galaxy.

\item At least for this simulation, we found that the final morphology of these systems is determined by $z \sim 1$, on average. 
It can be affected by  posterior events but, on average,  their morphology tends to be  dominated by previous processes. 
Gas accretion is required to build up the discs, but our simulation showed that galaxies that reach higher $D/T$ ratios are those that reach $z \sim 1$ already with a large disc component. 

\end{itemize}

Our results show that minor mergers might play a role in determinating the galaxy morphology by contributing to the formation of the disc components if they are aligned, or to the central spheroid, if they are misaligned, with respect to the main disc galaxy. This result  extends previous findings \citep{naab2009} by showing the role played by the alignment of the  minor systems in the formation of each dynamical component from $z \sim 2$. We detected for the first time that the simulated central spheroids formed from material which, although it has low angular momentum content, conserves a relation to that of the dark matter haloes, which might be relevant for understanding the mass-size evolution of galaxies.
We expect to go beyond this analysis and achieve a deeper understanding of the processes that determine the morphology with forthcoming
simulations with higher resolution and improved physics.

\begin{acknowledgements}
We thank the referee for useful comments that helped to clarify the manuscript. This work was partially supported by PICT 2011-0959 from ANPCyT, PIP 2009-0305 and PIP 2012-0396 awarded by CONICET from Argentina. PBT thanks support from the Millennium Institute of Astrophysics (MAS) and the Regular Grant UNAB 2014. Simulations were run in the Fenix cluster of the Numerical Astrophysics Group at the Institute for Astronomy and Space Physics.
\end{acknowledgements}

\bibliographystyle{aa}
\bibliography{disk_NEW_7_AA}

\end{document}